\documentclass[aps,twocolumn,showpacs]{revtex4}

\begin{document}

\title{Co-existence of Gravity and Antigravity:\\
The Unification of Dark Matter and Dark Energy}

\author{Xiang-Song Chen}
\affiliation{Department of Physics, Sichuan University,
                Chengdu 610064, China\\
        and Joint Center for Particle, Nuclear Physics and Cosmology, 
                Nanjing University, Nanjing 210093, China\\
{\rm Electronic address: cxs@scu.edu.cn, cxs@chenwang.nju.edu.cn}}
\date{\today}
                                                                            
\begin{abstract}
Massive gravity with second and fourth derivatives is shown to give both attractive and repulsive gravities.
In contrast to the attractive gravity correlated with the energy-momentum tensor, the repulsive gravity is related to a fixed mass $m_x$, which equals a spin-dependent factor $f_\sigma$ times the graviton mass. Therefore, particles with energy below $m_x$ are both dark matter and dark energy: Their overall gravity is attractive with normal matter but repulsive among themselves. Detailed analyses reveal that this unified dark scenario can properly account for the observed dark matter/energy phenomena: galaxy rotation curves, transition from early cosmic deceleration to recent acceleration; and naturally overcome other dark scenarios' difficulties: the substructure and cuspy core problems, the difference of dark halo distributions in galaxies and clusters, and the cosmic coincidence. Very interestingly, Dirac particles have $f_\sigma=1/\sqrt 2$, all bosonic matter particles have $f_\sigma=0$, and the only exceptional boson is the graviton itself, which may have $f_\sigma>1$.
\end{abstract}
\maketitle

In this rapid communication we show that the theoretical complications inherently contained in the gravitational interaction provide a natural and perfect solution to the cosmological dark matter/energy puzzles \cite{Sahni04}. 
If constructed as a quantum theory of a tensor field, gravity is intrinsically more complicated than the standard-model interactions both in the gravity-matter coupling and in the Lagrangian of the gravity itself. 
The dimensionlessness of the metric field allows for fourth-order derivatives without violating the power-counting renormalizability. The gravity-Dirac coupling, put in the standard tetrad formalism, also exhibits a rich structure: 
\begin{equation}\label{tetrad}
L=-\bar\psi \gamma^c \varepsilon _c\,^\mu(\partial_\mu+
\frac 12 \sigma^{ab} \omega_{\mu ab} )\psi,
\end{equation}
where $\varepsilon _c\,^\mu$ is the tetrad field, 
$\sigma^{ab}=\frac 14 (\gamma^a\gamma^b -\gamma^b\gamma^a)$, and 
\begin{equation}\label{omega}
\omega_{\mu ab}=\varepsilon_a\,^\nu(\partial_\mu \varepsilon_{b\nu}
-\partial_\nu\varepsilon_{b\mu})
+\frac 12 \varepsilon_a\,^\rho \varepsilon _b\, ^\sigma 
(\partial_\sigma \varepsilon_{c\rho}-\partial_\rho \varepsilon_{c\sigma})
\varepsilon^c\,_\mu.
\end{equation}

Unlike the bosonic case, Eq. (\ref{tetrad}) contains a derivative coupling to the tetrad field. As will be detailed below, although this derivative coupling is trivial in the standard Einstein gravity with massless graviton and only second derivatives, it becomes crucial if one considers the full theoretical complications allowed for gravity by including fourth derivatives and graviton mass. In fact, the one-graviton-exchange potential between Dirac particles is found to contain both attractive and repulsive terms with distinct properties. 

To explore the one-graviton-exchange potential, we write the tetrad field as $ \varepsilon ^a\,_\mu=\delta ^a\,_\mu+h^a\,_\mu +o(h)$, 
$ \varepsilon _a\,^\mu=\delta _a\,^\mu-h^\mu\,_a +o(h)$, and single out the terms linear in $h$ in Eq. (\ref{tetrad}). After dropping the total derivatives and making use of the equation of motion for the Dirac field, we get
\begin{equation}\label{linear}
L_1=h_{\mu\nu} \bar\psi \gamma^\mu \partial^\nu \psi +
\frac 12 \bar\psi \gamma^\mu\psi \partial^\nu h_{\mu\nu},
\end{equation}
where we have discarded the antisymmetric piece in $h_{\mu\nu}$ since the lowest order graviton propagator is symmetric:
\begin{equation}\label{prop}
\Delta(q^2)_{\mu\nu,\alpha\beta}=
\Delta(q^2)(\eta_{\mu\alpha}\eta_{\nu\beta}+\eta_{\mu\beta}\eta_{\nu\alpha}
-\eta_{\mu\nu}\eta_{\alpha\beta}).
\end{equation}
  
In massless Einstein gravity, $\Delta(q^2)\sim 1/q^2$, the derivative coupling in Eq. (\ref{linear}) just gives a trivial contact interaction. If we include fourth-order derivatives together with a mass term, the graviton propagator takes the form \cite{note1} 
\begin{equation}\label{fourth}
\Delta(q^2)\sim 1/(q^2+m_g^2)(q^2+M_{Pl}^2). 
\end{equation}

The coupling of Eq. (\ref{linear}) and the propagator of Eq. (\ref{fourth}) give a very interesting one-graviton-exchange potential: 
\begin{eqnarray}
V(r)\sim&-&\frac 1r \frac{e^{-m_g r}-e^{-M_{Pl} r}}{M_{Pl}^2-m_g^2}
\frac{(2P_1\cdot P_2)^2-m_1^2 m_2^2}{E_1E_2} 
\nonumber \\
&+&\frac 1r \frac {m_g^2 e^{-m_g r} -M_{Pl}^2 e^{-M_{Pl} r}}{M_{Pl}^2-m_g^2} \frac{P_1\cdot P_2}{2E_1E_2} ,
\label{potential}
\end{eqnarray}
where $P_{1,2}$ are the four-momenta of the two Dirac particles, $m_{1,2}$ are their rest masses, and $E_{1,2}$ are their relativistic energies. Except at very early universe with extremely high density, $e^{-M_{Pl} r}$ can be safely neglected.  
Eq. (\ref{potential}) clearly exhibits a standard attractive term correlated with energy-momenta of the two Dirac particles, and {\em a novel repulsive term proportional to the graviton mass $m_g$} \cite{chen05}, which tells that:

\begin{enumerate}
\item The repulsive term is negligible for normal matter Dirac particles (like quarks and electrons) whose masses are many orders of magnitude larger than $m_g$. 
\item For Dirac particles with energy comparable to $m_g$, the repulsive term becomes significant, and can dominate over the attractive term if, roughly speaking, the product of the two particles' energies is smaller than $m_g^2/2$. 
\item The repulsive gravity between normal matter and Dirac particles with energy comparable to $m_g$, on the other hand, is still negligible. 
\end{enumerate}

Such co-existence of attractive and repulsive gravities strongly violates the equivalence principle for Dirac particles with energy comparable to or much smaller than the graviton mass. It should be noted that bosonic matter fields have no derivative coupling to the graviton field, in consequence the gravity between bosons or between a Dirac particle and a boson does not contain the repulsive term. In the language we used in the Abstract, bosons have $f_\sigma=0$, and Eq. (\ref{potential}) says that Dirac particles have $f_\sigma=1/\sqrt 2$. However, there is one kind of boson which is exceptional: the graviton itself, which contains derivatives in its self-coupling. The $f_\sigma$ factor for the graviton strongly depends on construction of the complete graviton Lagrangian, which we postpone till later studies. At the moment we just assume that the graviton Lagrangian is constructed to give $f_\sigma >1$, so that at low enough energies the repulsive gravity can still dominate. 

The co-existence of attractive and repulsive gravities has far-reaching implications for cosmology. In what follows, we explain that antigravitating low-energy particles (ALEPs) can serve as both dark matter and dark energy. The ALEP paradigm is capable of solving all the observed cosmological dark matter/energy puzzles, and also naturally overcomes the difficulties associated with traditional dark scenarios. 

{\bf Dark halo formation and galaxy rotation curve.} 
ALEPs with energy lower than $m_x=f_\sigma m_g$ repel each other and do not cluster by themselves. However, they are attracted by normal matter and can condense around galaxies and form a dark halo. As the dark halo gets larger and larger, the attraction of the normal matter core is compensated and finally no more ALEPs can be attracted. On the other hand, this ALEP dark halo provides extra attraction for normal matter particles and contribute to the galaxy rotation curve.  

{\bf The dark halo substructure and cuspy core problem.} In traditional paradigms, the dark matter has the same gravitational property as the normal matter, which predicts too many dark matter sub-halos and too high dark matter density at the galaxy center. These are usually referred to as the substructure and cuspy core problems \cite{Sahni04}. The ALEP paradigm naturally solve these problems, since ALEPs can only condense around a normal matter core or when they have relatively high energies. This significantly reduces (but does not exclude) the probability of forming substructures. The repulsive gravity also prevents ALEPs from over-condensing, such as at the galaxy center, and makes the ALEPs more loosely bounded in comparison to the normal matter, therefore centrifugal force helps further to repel ALEPs from the galaxy center.     

{\bf Difference of dark matter distributions in galaxies and clusters.} Weak-lensing observations reveal that galaxy clusters contain more dark matter in the center than galaxies do. The ALEP paradigm provides two possible mechanisms for this phenomenon: (1) The centrifugal effect: galaxy clusters rotate much slower than galaxies do. (2) The acceleration effect: clusters have deeper gravitational potential well, thus when ALEPs are absorbed by clusters more of them can increase their energy to become condensable particles.

{\bf Early cosmic deceleration and recent acceleration:} Just like that heavier liquid falls closer to the bottom in a container, higher-energy ALEPs have higher priority to be attracted by the normal matter core, and lower-energy ones stay more to the outside. When the growth of the dark halo completes and no more ALEPs can be attracted, the relic lower-energy background ALEPs feel an overall repulsive gravity from the galaxy or cluster together with their dark halos. This in turn means that the background ALEPs provide a repulsive force for the halo-filled galaxy or cluster. When this background repulsion dominates over the attraction between the galaxies themselves, the cosmic expansion starts to speed up. Before the dark halo gets large enough, however, the ALEP background serves as attractive gravity source for the normal matter, and the cosmic expansion slows down. 

{\bf Cosmic coincidence.} The fine-tuning or cosmic coincidence problem is a major difficulty for many theories of cosmic acceleration, e.g., with a cosmological constant. This difficulty is also naturally avoided in the ALEP paradigm, which assigns only matter content to the universe and no cosmological constant or vacuum energy, yet the cosmic expansion inherently transits from deceleration to acceleration with the process of forming the ALEP dark halo.  

In closing, we remark that: 

(1) The ALEPs must have high enough density to be cosmologically relevant. Pauli exclusion principle dictates that the maximum number density of fermions with energy below the graviton mass is roughly one per cubic graviton Compton length, which is too low given any realistic graviton mass. Graviton density is not subjected to such limitation. 

(2) Many authors have suggested that dark matter and dark energy may come from the same source \cite{Bertolami05, Mainini05, Bento04, Mainini04, Khuri03, Bassett03, Padmanabhan02, Bilic02}. In most of these suggestions the dark components are of the traditional features. Namely, the dark matter has the same gravitational property as the normal matter and the dark energy repels galaxies directly in cosmic expansion. In our paradigm, however, ALEPS serve as both dark matter and dark energy because they attract the normal matter but repel each other. The repulsive effect overcomes several difficulties with the traditional dark matter scenarios. And in cosmic expansion, the repulsive force from the background ALEPs acts on the dark halo, which then pulls the normal matter core to accelerate together. The only author who explained an idea similar to ours about cosmic acceleration seems to be R. Khuri \cite{Khuri03}.  

(3) In this paper we have followed the deductive way of research. Namely, we first theoretically derived a repulsive gravity proportional to a fixed (graviton) mass, then found it fascinatingly capable of solving the dark matter/energy puzzles. Our analyses of these puzzles also suggest that {\em phenomenologically} the best solution to them is to assume the co-existence of the usual attractive gravity correlated with particle's energy-momentum and a novel repulsive gravity correlated with a fixed mass. By fitting the measurements one may figure out the magnitude of this fixed mass, and whether the repulsive gravity should be universal or only apply to certain type of particles.

{\it Acknowledgements:} I thank Terry Goldman for very helpful discussions. I also thank Terry and the T-16 group at Los Alamos National Lab, and Rupert Machleidt and the Physics Department at University of Idaho, for hospitality during my visit, during which part of this work was done. The work was supported by China NSF grant 10475057.


\begin{thebibliography}{99}
\bibitem{note1} We postpone till later studies the task of constructing an explicit Lagrangian, which is not uniquely determined by the lowest order propagator. In general, fourth-order derivative theories contain ghost states with negative norm. However, if properly formulated, the ghost states can be avoided or made irrelevant for physical process \cite{Hawking}.
\end{thebibliography}
\end{document}